\documentclass[aps,pra,twocolumn,nofootinbib,longbibliography,amssymb]{revtex4-1}
\usepackage{amsmath}
\usepackage{braket}
\usepackage{graphicx}
\usepackage[shortlabels]{enumitem}
\usepackage{substr}
\usepackage{verbatim}
\usepackage{dsfont}  \usepackage{mathtools}  \allowdisplaybreaks

\begin{document}

\title{A simple proof of Floquet's theorem for the case of unitary evolution in quantum mechanics}
\author{J.~D.~D.~Martin}
\author{A.~N.~Poertner}
\affiliation{Department of Physics and Astronomy, University of Waterloo, Waterloo, Canada}

\date{\today}

\begin{abstract}
We present a constructive proof of Floquet's theorem for the special case of unitary time evolution in quantum mechanics.  The proof is straightforward and suitable for study in courses on quantum mechanics.
\end{abstract}


\maketitle

Floquet's theorem \cite{floquet, isbn:978-0-471-86003-7} tells us something useful about the solutions to the time-dependent Schr\"{o}dinger equation (TDSE) for Hamiltonians that are periodic in time \cite{shortdoi:czxdv8}.  Specifically, given the TDSE:
\begin{equation}
i\hbar \frac{d \psi(t)}{d t} = H(t) \psi(t)
\end{equation}
in a space of finite dimension $n$,
with both $H(t)=H^{\dagger}(t)$ and $H(t+T)=H(t)$ for all $t$ with period $T$, there are $n$ orthonormal ``Floquet'' solutions of the form:
\begin{equation}
\psi_j(t) = \phi_j(t) e^{-i\epsilon_j t/\hbar}
\end{equation}
where $\phi_j(t) = \phi_j(t+T)$ for all $t$, and the $\epsilon_j$'s are real, so-called \emph{quasi-energies}.
For convenience here and in what follows, we consider $H(t)$, $U(t,0)$, etc... to be $n \times n$ matrix representations in an orthonormal basis; and likewise $\psi(t)$, $\phi_j(t)$, etc... to be $n \times 1$ column vectors.

The Floquet solutions may be used to construct a unitary time evolution operator:
\begin{equation}\label{eq:u_by_super}
U(t,0) = \sum_{j=1}^{n} \phi_j(t) e^{-i\epsilon_j t/\hbar} \phi_j^{\dagger}(0)
\end{equation}
appropriate for any initial conditions at $t=0$:
\begin{equation}
\psi(t) = U(t,0) \psi(0).
\end{equation}
Equivalently, we may form the $n \times n$ matrix
$\Phi(t) = [ \phi_1(t) \: \phi_2(t) \: \ldots \: \phi_n(t) ]$, and condense
Eq.~\ref{eq:u_by_super} to
\begin{equation} \label{eq:fundamental_result}
U(t,0) = \Phi(t) e^{-iE t/\hbar} \Phi^{\dagger}(0),
\end{equation}
where $E$ is a diagonal matrix containing the $\epsilon_j$'s, and $\Phi(t)$ is periodic; i.e., $\Phi(t) = \Phi(t+T)$ for all $t$.

Despite the relative simplicity of this result, it is not standard quantum mechanics textbook fare (with at least one exception \cite{isbn:9781891389238}).  But due to the fundamental nature of this ``time version'' of Bloch's theorem and recent activity in time-periodic quantum-mechanical Floquet systems  (see, for example, Ref. \cite{shortdoi:gfw8sp}), it may be desirable to study it in quantum mechanics courses.

The purpose of this short note is to present a simple, constructive proof of Floquet's theorem in the special case of quantum-mechanical unitary time evolution (Eq.~\ref{eq:fundamental_result}), appropriate for physics undergraduate students.  This proof arises naturally from consideration of how one might numerically compute time-evolution in an efficient manner.  The steps in the derivation are easily remembered, and involve only the standard results of basic quantum mechanics, as found in typical textbooks \cite{isbn:978-0-306-44790-7,isbn:9781108473224,isbn:978-1-107-18963-8}.

First note that for any positive integer $N$, the evolution from $t=NT$ to $t=(N+1)T$ as given by $U((N+1)T, NT)$ is the same as for from $t=0$ to $t=T$ as given by $U(T,0)$.  This symmetry can be seen by replacing $t$ by $t-T$ in the TDSE, since $H(t+T)=H(t)$ for all $t$.  Since $t$ in $U(t,0)$ is potentially much greater than $T$, it makes sense to compute $U(T,0)$, then raise it to the largest non-negative integer power $N$ such that $NT \le t$.
Then finally, compute evolution over the ``left-over'' interval from $NT$ to $t$ (of duration less than $T$).  In other words, we consider the factorization:
\begin{align}
 U(t,0) &= U(t,NT) \: U(NT, 0) \nonumber\\
        &=  U(t,NT) \: U(T, 0)^N. \label{eq:for_first_factor}
\end{align}
Taking the matrix power will be more computationally efficient if we first diagonalize $U(T,0)$.  Recall that any unitary operator has a complete orthonormal eigenbasis with complex eigenvalues of unit magnitude, a result that will be familiar to undergraduates (e.g., pg 39 of Ref.~\cite{isbn:978-0-306-44790-7}).  Thus we may write $U(T,0) = W e^{iD} W^{\dagger}$, where $D$ is a real, diagonal, $n \times n$ matrix, and $W$ is unitary ($W^{\dagger}W=W W^{\dagger}= \mathds{1}$).
This diagonalized form of $U(T,0)$ is more efficient for computation because
\begin{align}
  U(T,0)^N &= (W e^{iD} W^{\dagger})^N \nonumber\\
           &= W e ^{iD} W^{\dagger} \: W e^{iD} W^{\dagger} \ldots
              W e^{iD} W^{\dagger} \nonumber\\
           &= W e^{iDN} W^{\dagger}.
\end{align}
As it is unlikely that $t$ is an exact integer multiple of $T$, we still need to compute time evolution over the remaining time interval from $NT$ to $t$ (the first ``factor'' on the RHS of Eq.~\ref{eq:for_first_factor}).  The time-periodic $H(t)$ allows us to write the remaining time evolution as:
\begin{equation}
  U(t,NT) = U(t-NT, 0).
\end{equation}
It is convenient to introduce a notation for the ``leftover'' time (after Ref.~\cite{isbn:978-0-201-55802-9}).  Defining $\lfloor x \rfloor$ as the largest integer less than or equal to $x$, we have $N = \lfloor t/T \rfloor$.  If we also define ``$\operatorname{mod}$'', so that $x \operatorname{mod} y \coloneqq x - y \lfloor x/y \rfloor$, then $t-NT = t \operatorname{mod} T$.  Note that $t \operatorname{mod} T$ is periodic in $t$, with period $T$.

Combining evolution from $0$ to $NT$ with evolution from $NT$ to $t$ gives:
\begin{align}
U(t, 0) &= U(t, NT) \:  U(NT, 0)\\
        &= U(t \operatorname{mod} T, 0) \: W e^{iDN} W^{\dagger}\\
\intertext{and using $N=[t- (t \operatorname{mod} T)] /T$ gives:}
U(t,0)  &= U(t \operatorname{mod} T, 0) \: W e^{iD[t-(t \operatorname{mod} T)]/T} W^{\dagger} \label{eq:almost_last}.
\end{align}
Defining $\Phi(t) \coloneqq U(t \operatorname{mod} T, 0) \: W e^{-iD(t \operatorname{mod} T)/T}$, we note that it is both periodic and unitary.  Also defining $E \coloneqq -D \hbar /T$, we note that it is real and diagonal.
Using these definitions we may rewrite Eq.~\ref{eq:almost_last} as
\begin{equation}
U(t,0) = \Phi(t) e^{-iEt/\hbar} \Phi^{\dagger}(0),
\end{equation}
the result that we were to show (Eq.~\ref{eq:fundamental_result}) $\square$.

Given the periodicity of $\Phi(t)$, it is natural to consider its Fourier expansion.  Shirley \cite{shortdoi:czxdv8} followed this line of thought, showing that both the Fourier expansion of $\Phi(t)$ and the corresponding quasi-energies may be obtained by diagonalization of a \emph{time-independent} Hamiltonian, constructed using the Fourier components of $H(t)$.

A preliminary version of this note appeared in the supplemental material for Ref.~\cite{shortdoi:ghcg7z}.

\bibliography{bib_files/floquet.bib,bib_files/bibtex_from_zotero.bib}

\IfSubStringInString{\detokenize{generated}}{\jobname}{
\verbatiminput{git_info_generated.txt}
}{}

\end{document}